\documentclass[reprint,amsmath,amssymb,aps,prb,showpacs]{revtex4-1}

\usepackage{graphicx}
\usepackage{epsfig}
\usepackage{dcolumn}
\usepackage{bm}
\usepackage{hyperref}
\hypersetup{backref=true,
 pdfnewwindow=true, colorlinks=true,
 linkcolor=blue, anchorcolor=blue,
 citecolor=blue, filecolor=blue,
 menucolor=blue, urlcolor=blue}

  \usepackage[usenames,dvipsnames]{color}

\begin{document}
\title{Smooth gauge for topological insulators}
\author{Alexey A. Soluyanov}
\email{alexeys@physics.rutgers.edu}
\author{David Vanderbilt}
%\email{dhv@physics.rutgers.edu}
\affiliation{Department of Physics and Astronomy, Rutgers University,
Piscataway, New Jersey 08854-0849, USA}
\date{\today}

\def\TR{{\cal T}}

\begin{abstract}
We develop a technique for constructing
Bloch-like functions for 2D $\mathbb{Z}_2$-insulators (i.e., quantum spin-Hall
insulators) that are smooth functions of ${\bf k}$
on the entire Brillouin-zone torus.  As the
initial step, the occupied subspace of the insulator is
decomposed into a direct sum of two ``Chern bands,'' i.e.,
topologically nontrivial subspaces with opposite
Chern numbers.  This decomposition
remains robust independent of underlying symmetries or specific
model features.
Starting with the Chern bands obtained in this way,
we construct a topologically
nontrivial unitary transformation that rotates the
occupied subspace into a direct sum of topologically
trivial subspaces,
thus facilitating a Wannier construction.
The procedure is validated and
illustrated by applying it to the Kane-Mele model.
\end{abstract}
\maketitle

% %%%%%%%%%%%%%%%%%%%%%%%%%%%%%%%%%%%
% \marginparwidth 2.7in
% \marginparsep 0.5in
% \def\dvm#1{\marginpar{\small DV: #1}}
% \def\asm#1{\marginpar{\small AS: #1}}
% \def\scr{\scriptsize}
% %%%%%%%%%%%%%%%%%%%%%%%%%%%%%%%%%%%

%%%%%%%%%%%%%%%%%%%%%%%%%%%%%%%%%%%
\marginparwidth 2.7in
\marginparsep 0.5in
\newcounter{comm} % counter for commentaries
% increase counter
\def\commnext{\stepcounter{comm}}
% commentary in text
\def\commtext{{\bf\color{blue}[\arabic{comm}]}}
% commentary in margin
\def\commmar{{\bf\color{blue}[\arabic{comm}]}}
% comment commands for all authors
\def\dvm#1{\commnext\marginpar{\small DV\commmar: #1}\commtext}
\def\asm#1{\commnext\marginpar{\small AS\commmar: #1}\commtext}
\def\tnewpage{\newpage\marginpar{\small Temporary newpage}}
\def\scr{\scriptsize}
%%%%%%%%%%%%%%%%%%%%%%%%%%%%%%%%%%%

\def\r{{\bf r}}
\def\R{{\bf R}}
\def\k{{\bf k}}
\def\G{{\bf G}}
\def\z2{{\mathbb{Z}_2}}

\section{Introduction}
In recent years the band theory of solids has been
augmented by new chapters to account for
geometric and topological effects that had not
been considered previously.  The introduction of the
Berry phase\cite{Berry-PRSL84} allowed the systematic
description of many observable effects of purely geometric
origin, such as the Aharonov-Bohm effect,\cite{Aharonov-PR59}
and its applications in the band-theory context have included the
theory of electric polarization\cite{King-Smith-PRB93,Resta-RMP94}
and the anomalous Hall conductance.\cite{Thouless-PRL82,Haldane-PRL04}

The recent discovery of topological
insulators\cite{Hasan-RMP10,Qi-RMP11} has widened the
role of geometry and topology in band theory even
further. The classification of non-interacting insulating
Hamiltonians in 2D predicts two topologically
nontrivial scenarios.\cite{Kitaev-AIP09,
Schnyder-PRB08,Schnyder-AIP09} The first scenario is that of
a Chern insulator, i.e., an insulator that exhibits an integer quantum
Hall effect even in the absence of an external magnetic
field.\cite{Haldane-PRL88} Such a material, also known as
a quantum anomalous Hall insulator, is classified according
to the value of transverse conductance in integer multiples of
$e^2/h$, i.e., a $\mathbb{Z}$ classification.
The $\mathbb{Z}$ invariant contains information about
the excess chirality of current-carrying edge states of a
2D sample.  Hamiltonians that correspond to different
integers represent distinct topological phases, meaning
that they cannot be adiabatically connected without closing
the insulating gap.\cite{Kitaev-AIP09,
Schnyder-PRB08,Schnyder-AIP09}
Chern insulators break time-reversal (TR)
symmetry, since $\sigma_{xy}$ is odd under TR.
The name arises from the fact that the exact
quantization of conductance is of topological origin, i.e.,
the conductance is written as
$\sigma_{xy}=C(e^2/h)$ where $C$ is called the Chern
number or TKNN invariant.\cite{Thouless-PRL82, Kohmoto-AP85, Nakahara-book}

The second scenario in 2D is that of a TR-symmetric $\mathbb{Z}_2$
insulator\cite{Kane-PRL05-b} that possesses either an odd or an even
number of Kramers pairs of edge states. According to the
number of these pairs at the edge, the insulator is either
$\z2$-odd or $\z2$-even. These two phases are topologically
distinct and cannot be adiabatically connected to one another
without gap closure. A $\z2$-odd insulator realizes the quantum spin Hall
(QSH)\cite{Kane-PRL05-b, Kane-PRL05-a} state, while a $\z2$-even
one is adiabatically connected to a normal insulator.
In what follows we sometimes refer to the QSH insulator
as a ``$\z2$ insulator.'' Unlike the Chern-insulator state,
the QSH-insulator state has been realized
experimentally, e.g., in CdTe/HgTe/CdTe quantum
wells\cite{Konig-Science07} following a theoretical
prediction.\cite{Bernevig-Science06,Volkov-JETP85}

\begin{figure}
\begin{center}
\includegraphics[width=2.8in]{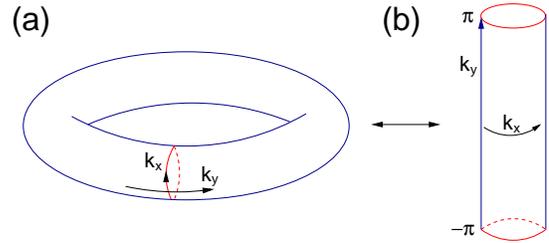}
\end{center}
\caption{Brillouin zone in 2D represented as (a) a torus, and (b) a
cylinder. We choose the gauge discontinuity to be distributed
along the cross-sectional cut of the torus that maps onto the
end loops of the cylinder at $k_y=\pm\pi$.}
\label{fig:1}
\end{figure}

On the level of conventional band theory of crystalline solids,
Chern and $\z2$ insulators are also different from ordinary
ones. For an ordinary insulator the Bloch states $\psi_{n\k}$
are usually assumed to be smooth and periodic in the Brillouin
zone (BZ), meaning that a translation by
a reciprocal lattice vector $\G$ returns the Bloch wavefunction
back to itself with the same phase, $\psi_{n,\k+\G}=
\psi_{n\k}$, and that $\psi$ is a smooth function of $\k$.
Regarding the BZ as a torus, as in Fig.~\ref{fig:1}(a), this just
means that $\psi$ is a smooth function of $\k$ on the torus.
This turns out to be impossible for Chern
insulators;\cite{Thouless-JPC84,Thonhauser-PRB06}
the occupied space of a Chern insulator cannot be represented by
smooth and periodic Bloch states. Usually periodicity is still
assumed, in which case a point discontinuity or branch cut must
appear in the phase of at least one occupied Bloch state somewhere in
the BZ. It is now established that no gauge transformation -- i.e.,
no $\k$-dependent unitary rotation of the bands in the occupied subspace
-- can smooth out this discontinuity.\cite{Thouless-JPC84,Thonhauser-PRB06}

In the case of $\z2$ insulators, the presence of TR symmetry
forces the total Chern number to vanish, guaranteeing the existence,
in principle, of a smooth and periodic gauge in the
BZ.\cite{Brouder-PRL07} However, it has been shown that any gauge
that respects TR symmetry cannot be smooth on the torus for this
class of topological materials.\cite{Fu-PRB06,Roy-PRB09-a,Loring-EPL10}
Thus, the construction has to break TR symmetry if it
is to lead to a smooth gauge.  An explicit construction of
this type for the QSH model of Kane and
Mele demonstrated that this is possible,\cite{Soluyanov-PRB11-a}
but the method used there was explicitly model-dependent, and it
remained unclear how one should choose a smooth gauge
for a generic $\z2$ insulator.

In the present paper we address this question and develop a general
procedure for constructing smooth and periodic Bloch states for
QSH insulators. We limit ourselves to the minimal case of two
occupied bands and show how they can be disentangled into two
single-band subspaces having equal and opposite Chern numbers,
in such a way that these subspaces are mapped
onto each other by the TR operator $\theta$.

Each of these ``Chern bands'' has the same type of
gauge discontinuity on the boundary as is present
in a Chern insulator.  The possibility of such a
decomposition has been discussed before in different
contexts,\cite{Roy-PRB09-a,Sheng-PRL06,Teo-PRB08,Prodan-PRB09}
but the previous approaches all have relied on some specific
feature of the system, such as separation of states according to
the action of the $S_z$ or mirror symmetry operators.
Instead, our construction is based on topological considerations
alone, and should remain robust for any $\z2$ insulator.
We further impose on these Chern bands a special ``cylindrical
gauge'' in which the gauge discontinuity is spread uniformly
around the circular cross section of the BZ torus, i.e.,
connecting the end loops at $k_y=\pm\pi$ in Fig.~\ref{fig:1}(b).
Finally, we develop a procedure that mixes these two topologically
nontrivial states in such a way that they become smooth and
periodic in the BZ, thus obtaining a smooth (but TR-broken)
gauge.

Apart from the purely theoretical motivation, the problem of
constructing smooth Bloch states for $\z2$ insulators has
a direct practical application. When working with ordinary
band insulators it is often convenient to use a real-space
formulation in terms of the Wannier representation.
In this representation, the occupied subspace is described by
a lattice of Wannier
functions that are exponentially localized in real space.
The Wannier representation is very useful for computing many
properties of insulating materials, such as electric polarization,
charge distributions, or bonding properties, or when constructing model
Hamiltonians.\cite{Marzari-PRB97,Vanderbilt-PRB93,Zak-PRL89,Marzari-RMP12}
However, exponentially localized Wannier functions may be
constructed only out of a set of smooth Bloch states. Thus,
construction of a smooth gauge for $\z2$ insulators allows
for the use of well-established Wannier-based methods in the study of
these materials.

Another interesting aspect of the present work arises
from the fact that a smooth gauge allows one to compute
the $\z2$ topological invariant directly by tracing
the connectivity of the states between some special
points in the BZ.\cite{Fu-PRB06} In the presence of
inversion symmetry this task is greatly simplified,\cite{Fu-PRB07}
since inversion symmetry allows one
to choose states that are smoothly connected in the BZ.
In the absence of inversion symmetry, however, the
same is not true,
and the computation of the topological invariant also becomes
more complicated.\cite{Fukui-JPSJ07,Soluyanov-PRB11-b,
Yu-PRB11, Prodan-PRB11} Thus, one can consider the present method
as an alternative recipe for computing topological
invariants.

The present work treats the two-dimensional case.  For a
three-dimensional TR-invariant insulator, the method described
here can be used to construct a smooth gauge on any of the
six TR-invariant planes in the BZ. However, the final
connection between these faces to obtain a globally smooth gauge
in 3D appears to be nontrivial except in special cases (e.g.,
certain kinds of weak topological insulators).  A general
formulation in 3D is therefore left to future investigations.

We emphasize that questions of gauge choice do not
affect physical observables such as the dispersions or spin
textures of the energy bands.  Thus, if used properly, even a gauge
that violates TR symmetry, or that has a gauge discontinuity on
the BZ boundary, should be capable of making robust predictions
of physical properties consistent with TR symmetry.
We are concerned here with formal issues of
gauge construction and practical questions about which construction
is most convenient for computing physical properties.

The paper is organized as follows. The specific
gauge that we want to establish and the concept
of individual Chern numbers are introduced in
Sec.~\ref{sec:cherns}. The procedure for
disentangling the occupied subspace of a $\z2$
insulator into Chern subspaces is described in
Sec.~\ref{sec:decomp}, where we also
discuss the relation of our decomposition
procedure to ones discussed elsewhere.
Sec.~\ref{sec:rotation} introduces a general
procedure for constructing a smooth gauge out of the two
Chern subspaces. We give our conclusions in
Sec.~\ref{sec:concl}. Finally, the paper includes
three appendices. In App.~\ref{app:partr} we
describe the parallel transport of states,
a procedure that is used heavily in the construction of
the Chern bands. App.~\ref{app:KMmodel} presents a brief
summary of the Kane-Mele model\cite{Kane-PRL05-b} that we use
to illustrate our method. Finally, the relation of the
smooth gauge constructed in the present paper to the one
discussed by Fu and Kane in Ref.~\onlinecite{Fu-PRB06} is
discussed in App.~\ref{app:trs}.

\section{Cylindrical gauge and individual Chern numbers}
\label{sec:cherns}
In this section we consider the definition of the
Chern number of a Bloch band in 2D and introduce a
cylindrical gauge for Chern bands. This is a gauge
that is continuous in the BZ but
is periodic in $k_x$ only.  That is, it is continuous
on the cylinder in Fig.~\ref{fig:1}(b), but not across the
boundary connecting top to bottom, i.e., not on the torus
of  Fig.~\ref{fig:1}(a).  We then establish the notion of individual
band Chern numbers in the multiband case.

%-----------------------------------------------------
\subsection{Single band case}
\label{sec:single}
%-----------------------------------------------------

Let us first consider a single isolated Bloch
band $\psi_{\k}(\r)$ in 2D and its
cell periodic part $u_{n{\bf k}}({\bf r})=e^{-i{\bf k}
\cdot{\bf r}}\psi_{\bf k}({\bf r})$.
We assume the lattice vectors to have unit length and to be aligned
with the Cartesian axes, i.e., ${\bf a}_1=\hat{x}$ and
${\bf a}_2=\hat{y}$, so that $k_x$ runs from 0 to $2\pi$ 
and $k_y$ runs from $-\pi$ to $\pi$.
(In the general case, a linear transformation trivially rescales
the $k$ indices into this form.)
The Abelian Berry connection\cite{Berry-PRSL84} associated with these Bloch functions
is introduced as
\begin{equation}
\bm{\mathcal{A}}({\bf k})=
    i\langle u_{\bf k}| \nabla{\bf k} | u_{\bf k} \rangle
\label{A1def}
\end{equation}
and the corresponding curvature becomes
\begin{equation}
{\cal F}=\nabla_{\bf k} \wedge \bm{\mathcal A}= -2\mathrm{Im}\langle
\partial_{k_x}u_{\bf k}|\partial_{k_y}u_{\bf k}\rangle.
\label{F1def}
\end{equation}
It is important to note that, unlike the Berry connection, the
curvature is a gauge-invariant quantity.  Since we are in 2D, the
BZ is represented by the torus $T^2$ shown in
Fig.~\ref{fig:1}(a), which is a closed manifold. The integral of the
Berry curvature over the closed manifold is
necessarily a multiple of $2\pi$, and the integer number
\begin{equation}
C
=\frac{1}{2\pi}\int_{BZ} d^2k{\cal F}({\bf k}),
\label{C1def}
\end{equation}
is called a Chern number.\cite{Nakahara-book}
In general, the non-zero Chern number
reflects the impossibility of constructing
a periodic gauge without the presence of points or lines in the BZ
where the wavefunction would have a phase discontinuity.

To have a particular example of a gauge that leads to a nonzero
Chern number $C$, consider a gauge that is smooth everywhere on
the BZ torus except on a circle as shown in Fig.~\ref{fig:1}(a).
Any gauge discontinuity that might be present has thus been pushed
to this circular boundary, where the phase of the wavefunction can
experience a jump when crossing it.  Such a gauge is continuous on
the cylinder formed by cutting the torus along the discontinuity,
shown in Fig.~\ref{fig:1}(b), but is not periodic in the $y$
direction.  We now define a ``cylindrical" gauge to be one in
which the gauge discontinuity is uniformly distributed around
the boundary.  That is, such a gauge obeys the boundary conditions
\begin{eqnarray}
\psi_{{\bf k}+2\pi\hat{x}}&=&\psi_{\bf k}, \nonumber\\
\psi_{{\bf k}+2\pi\hat{y}}&=&\psi_{\bf k}\,e^{iCk_x},
\label{1band2Dpsi}
\end{eqnarray}
or, equivalently,
\begin{eqnarray}
u_{{\bf k}+2\pi\hat{x}}&=&e^{-2\pi ix}\,u_{\bf k}, \nonumber\\
u_{{\bf k}+2\pi\hat{y}}&=&e^{-2\pi iy}\,u_{\bf k}\,e^{iCk_x},
\label{1band2D}
\end{eqnarray}
where $C$ is the Chern integer.
The cylindrical gauge is assumed to be continuous
inside the rectangle of the BZ and $G_x$-periodic
in $k_x$, so it is continuous on the cylinder.
This leads to the continuity of the vector field
$\bm{\mathcal A}({\bf k})$
on the cylinder and, hence, Gauss's theorem may be applied to
the definition~(\ref{C1def}) to write
\begin{equation}
C=\frac{1}{2\pi}\oint_{\partial{\rm BZ}}
\bm{\mathcal A}({\bf k})\cdot d{\bf k},
\label{C1A2D}
\end{equation}
where the boundary $\partial{\rm BZ}$ of the BZ
consists of the top and bottom loops ($S^1\oplus S^1$) of the
cylinder at $k_x=-\pi$ and $\pi$.
From Eq.~(\ref{C1A2D})
the consistency of the chosen gauge
with the definition of the Chern number
becomes obvious. That is, $C$ in the exponent of
the boundary conditions of Eqs.~(\ref{1band2Dpsi}-\ref{1band2D})
is exactly the Chern number. Note that
since we consider here a single isolated band,
this Chern number is a gauge-invariant quantity.

%---------------------------------------------------------------------------
%\subsection{Individual Chern numbers}
\subsection{Multiband case and individual Chern numbers}
%-----------------------------------------------------------------------

Let us now consider ${\cal N}>1$ bands separated
by energy gaps from the rest of the spectrum.  The
Abelian connection of Eq.~(\ref{A1def}) is now replaced by its
non-Abelian multiband generalization\cite{Wilczek-PRL84,
Mead-RMP92}
\begin{equation}
{\cal A}_{mn,\alpha}=
i\langle u_{m{\bf k}}|\partial_{\alpha}|u_{n{\bf k}}\rangle
\label{A2def}
\end{equation}
and the non-Abelian curvature is defined as
\begin{equation}
F_{mn,\alpha \beta} =
{\cal F}_{mn, \alpha \beta}-i[{\cal A}_\alpha, {\cal A}_\beta]_{mn},
\end{equation}
where the ${\bf k}$-dependence is implicit.  $F$ is gauge-covariant
and $\mathrm{Tr}[F]$ is gauge-invariant\cite{Nakahara-book} under
a general unitary transformation ${\cal U}\in{\mathrm{U}({\cal N})}$ of
the occupied bands, i.e.,
\begin{equation}
|u_{n\k}\rangle = \sum_j {\cal U}_{jn}(\k)|u_{j\k}\rangle.
\label{gtrans}
\end{equation}
The Chern number is now assigned to the entire space of ${\cal N}$ bands
and is defined as
\begin{equation}
C=\frac{1}{2\pi}\int_{BZ}d^2k \mathrm{Tr}[F]=
\frac{1}{2\pi}\int_{BZ}d^2k \mathrm{Tr}[{\cal F}],
\label{ctot}
\end{equation}
where the trace is taken over the band index.

If we now suppose that in the group of bands under consideration
each of the ${\cal N}$ bands is isolated -- that is, separated from
the others by finite gaps -- then the total Chern number of
the subspace is just the sum
\begin{equation}
C=\sum_{n=1}^{\cal N} c_n
\label{ccconstr}
\end{equation}
of the individual Chern numbers of all the bands in the subspace,\cite{Avron-PRL83}
where $c_n$ are computed for isolated bands as described
in the preceding section. Being treated in this way, each $c_n$
is an integer. However, one might be tempted to define the quantity
\begin{equation}
\tilde{c}_n=\frac{1}{2\pi}\int_{BZ}d^2 k\,{\cal F}_{nn,xy}({\bf k})
\label{indChern}
\end{equation}
as the single-band contribution of band $n$ to $C$. Thus defined,
$\tilde{c}_n$ is {\it not} necessarily an integer, since it
is now allowed to mix the bands by a transformation of
the form of Eq.~(\ref{gtrans}), which can change ${\cal F}_{nn,xy}$.
Hence, in the multiband case the partial Chern contributions
defined by Eq.~(\ref{indChern}) are not topologically
invariant.

The example of a group of isolated bands suggests that in certain
gauges the subspace under consideration may be decomposed
into the direct sum of smaller subspaces for which Chern numbers
are well defined. In this particular example the gauge that
naturally realizes this decomposition is the Hamiltonian gauge,
that is, the gauge in which the Hamiltonian is diagonal.
However, one might wonder whether such a decomposition is still
possible for overlapping bands.

%\dvm{I'm not so sure about this paragraph.  It seems to step off
%the main line of argument and is a little fuzzy.  Should we
%consider dropping it?}
%
%This question does not arise for a topologically trivial
%band structure unless one looks for the smoothest possible
%bands,\cite{Souza-PRB01} since when all the bands that overlap
%have $c_n=0$, it is possible to construct different superpositions
%of these bands that will remain topologically trivial.  On the
%other hand, when mixing bands with non-zero Chern numbers, the
%whole notion of an {\it integer} topological invariant for the
%individual band $c_n$ is lost, and one is left with the intuitive
%definition of Eq.~(\ref{indChern}). To give physical sense to this
%formula, one has to fix a specific gauge such that the $\tilde{c}_n$
%become integers.

A QSH insulator has a nontrivial topology,\cite{Kane-PRL05-b}
which can be seen as an obstruction for constructing smooth
Bloch functions in a gauge that respects the TR symmetry
of the Hamiltonian.\cite{Fu-PRB06,Roy-PRB09-a,Loring-EPL10}
In what follows we describe a generic procedure for decomposing
the occupied subspace of such an insulator into a direct sum of
Chern subspaces, i.e., disentangling
the occupied subspace into bands with well-defined
individual integer Chern numbers $c_n$. We also show that
each of these Chern bands may be represented in the cylindrical
gauge of Eq.~(\ref{1band2D}) with $C$ replaced with ${c}_n$.

\section{Decomposition into Chern subspaces}
\label{sec:decomp}

In this section we develop a general procedure for disentangling
a Kramers pair of occupied states of a 2D $\z2$-insulator into two
Chern bands with individual Chern numbers $c_1=-1$ and $c_2=1$. The
decomposition method makes heavy use of the concept of parallel transport
described in Appendix~\ref{app:partr}. The procedure is illustrated
by its
application to the Kane-Mele model that is reviewed in
Appendix~\ref{app:KMmodel}. We start by using parallel transport
of the Bloch states to move the gauge discontinuity to the edge of the BZ.
This makes the gauge continuous on the cylinder in $k$-space.
The next step is to apply certain gauge transformations to
split the occupied subspace into a direct sum of two subspaces
that are mapped onto one another by TR. We then explain
how to impose the cylindrical gauge on the two
disentangled bands. Since by the time of this step the
bands are already continuous in the
interior of the cylinder, it is only the form of the
discontinuity at the edge
that has to be modified. Finally, we discuss the relation of our
decomposition to the previously proposed ``spin Chern numbers.''

\subsection{Moving the gauge discontinuity to the BZ edge}
\label{subs:movg}
%
%\dvm{On a second reading, it seemed to me that much of what
%was your first paragraph of Subsec.~A was redundant with what
%was already said in the preceding paragraph.  So I condensed
%this and reorganized with the next paragraph.  I also tried to keep
%the formalism open to apply also to a normal insulator, specializing
%to the $\z2$-odd case a bit later.}
%
We now consider a general model of a TR-symmetric
insulator in 2D. For simplicity we consider a minimal model with
two occupied bands only,
since it is the Kramers pairs near the Fermi level that are
responsible for a topological phase.
%
%\dvm{A general comment about the paper: A referee may say ``Well,
%they did everything for two occupied bands.  Then in one
%sentence they wave their hands about the case of more than two
%occupied bands.  Can the procedure really be generalized to the
%multiband case of $2N$ occupied bands?"  Do we want to say more
%about this question?  Maybe not here, but maybe at the end of
%III.B and/or III.C, sketching briefly how the procedure would
%be generalized to more bands?  Or near the very end?}
%
Thus, we consider
the solution of the Schrodinger equation
$H(\k)\vert u_{n\k}\rangle =E_{n\k}\vert u_{n\k}\rangle$
under the TR-invariance condition
$\theta H(\k) \theta^{-1}=H(-\k)$.
As was discussed above, the BZ is assumed to have been reduced
to a square spanning $[0,2\pi]\times[-\pi,\pi]$.

We start by taking two occupied states $|u_1\rangle$ and $|u_2\rangle$
resulting from numerical diagonalization at $(0,0)$. By TR
invariance, these must be Kramers-degenerate at this point.
Numerical diagonalization brings random phases to both states;
we accept the random phase assigned to $|u_1\rangle$, but ensure
that the second state is a Kramers partner to the first by setting
$|u_2\rangle=\theta |u_1\rangle$.
Starting from these states we move the gauge
discontinuity to the edge of the BZ in several steps.

\paragraph*{Parallel transport along $k_x$ at $k_y=0$.}

As a first step of our procedure, we carry out a multiband parallel
transport from $\k=(0,0)$ to $(2\pi,0)$ along the $k_x$ axis.
This procedure is described in detail in Appendix~\ref{app:partr},
but in brief it works as follows.
Starting from the the two occupied states at $(0,0)$, we step
along a mesh of $k_x$ values, each time carrying out a $2\times2$
unitary rotation of the two states at the new $k_x$ such that
the $2\times2$ matrix of overlaps with the states at the previous
$k_x$ is as close as possible to the identity.  The $2\times2$
unitary matrix $U$
relating the states $\psi_{n\k}$ at $(2\pi,0)$ to those at at
$(0,0)$ (i.e., the $\Lambda$ matrix of Eq.~(\ref{lambda}))
is then constructed; its
eigenvalues $\lambda_n=e^{i\phi_n}$ yield the non-Abelian Berry
phases $\phi_n$.\cite{Wilczek-PRL84, Mead-RMP92}   In the
present case, the TR symmetry ensures that $\lambda_1=\lambda_2$,
so that $U$ is just the identity times $e^{i\phi}$ where
$\phi=\phi_1=\phi_2$.  Finally, the gauge discontinuity from
$(2\pi,0)$ back to zero is ``ironed out'' by applying the gradual
phase rotation $e^{-i\phi k_x/2\pi}$ to the two states at each $k_x$.

\begin{figure*}
\begin{center}
\includegraphics[width=6.0in]{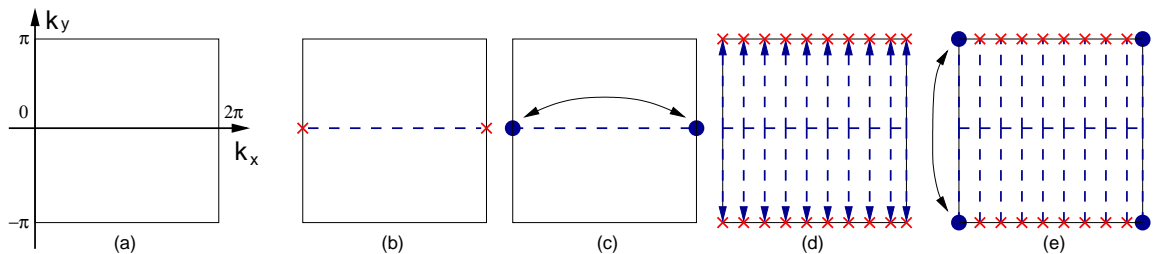}
\end{center}
\caption{(a) BZ in $k$-space. (b) States are parallel
transported along $k_y=0$, but are not periodic.
(c) Periodicity is restored at $k_y=0$. (d) Parallel
transport of states at all $k_x$ from $k_y=0$ to
$k_y=\pm\pi$. (d) Periodicity is restored at $k_x=0$,
and, hence, $k_x=2\pi$, but not at other $k_x$.}
\label{fig:2}
\end{figure*}

As a result of this procedure, we have a set of states that
are smooth functions of $k_x$ on the circular cross section
of the BZ torus at $k_y=0$,
including across the seam connecting $k_x$=0 to $k_x=2\pi$.
This is illustrated schematically in Fig.~\ref{fig:2}(b-c).
\paragraph*{Parallel transport along $k_y$ at each $k_x$.}

Next, at each mesh point $k_x$, we carry out two independent
parallel-transport procedures, one from $(k_x,0)$ to $(k_x,\pi)$
along $+\hat{y}$ and another from $(k_x,0)$ to $(k_x,-\pi)$ along
$-\hat{y}$. At each new $k_y$ point, the states are rotated by a
unitary matrix so that the matrix of overlaps with the previous
pair is as close to unity as possible.
Starting this procedure from the line $k_y=0$ guarantees that the
states on this line remain unchanged, preserving the smoothness
obtained previously.
Moreover, the entire parallel-transport procedure is identical
at $k_x=0$ and $k_x=2\pi$, ensuring that the states selected in
this way are continuous across the entire seam where  $k_x=0$ has
been glued to $k_x=2\pi$.
Thus,  we end up with two
states defined everywhere on the mesh of $k$ points in such a way
that they are smooth inside the BZ and periodic in $k_x$, or
equivalently, smooth everywhere on the cylinder of
Fig.~\ref{fig:1}(b).  This step is illustrated in Fig.~\ref{fig:2}(d).
The above procedure relates the states at $(k_x,-\pi)$ to those
at $(k_x,\pi)$ by a unitary matrix
\begin{equation}
V_{mn}(k_x)=\langle u_{m(k_x,k_y=-\pi)}|
\, {e^{2\pi iy}} \,
|u_{n(k_x,k_y=\pi)} \rangle ,
\label{Vdef}
\end{equation}
which plays a role similar to $\Lambda$ of Eq.~(\ref{lambda}).
This matrix encodes the information about the gauge discontinuity
that occurs on the boundary of the cylindrical BZ.
Its off-diagonal elements contain information about
entanglement of the two states, while the diagonal
ones carry information about phase discontinuity of
the states.

\paragraph*{Restoring periodicity in $k_y$ at $k_x=0$.}

The fact that the two states at ${\k}=0$ form a Kramers pair
guarantees that the matrix $V(k_x)$ is diagonal at $k_x=0$ with two
degenerate eigenvalues $\lambda(k_x=0)= e^{i\varphi_0}$.
(Incidentally, the same is true at $k_x=\pi$; we use this
fact later.)
Now we want to restore the smoothness across $k_y=\pm\pi$ at
$k_x=0$, but in such a way as to preserve the smoothness inside
the cylindrical BZ. We do this by multiplying all states by a
phase factor that depends smoothly on $k_y$ only:
\begin{equation}
|u_{n(k_x,k_y)}^{\rm new}\rangle = e^{-ik_y\varphi_0
/2\pi }
|u_{n(k_x,k_y)}\rangle
\end{equation}
After this transformation, the $V(k_x)$ matrix is the
identity at $k_x$=0.  Thus, the gauge discontinuity, which
has already been segregated to the edges at $k_y=\pm\pi$,
has now been further excluded from the point lying at $k_x$=0
(or $2\pi)$ on the edge.
Fig.~\ref{fig:2}(e) illustrates this, where red crosses
on the edges represent the gauge discontinuity and the black dots
indicate continuity.

Note that the entire procedure up to this point preserves the
TR symmetry, so that the states obtained so far on the
BZ respect the constraints
\begin{eqnarray}
\theta|u_{1{\bf k}}\rangle&=&|u_{2-{\bf k}}\rangle, \nonumber\\
\theta|u_{2{\bf k}}\rangle&=&-|u_{1-{\bf k}}\rangle.
\label{chi0}
\end{eqnarray}
This in turn implies that
\begin{equation}
V(-k_x)=\sigma_y \, [V(k_x)]^T \, \sigma_y
\label{VV}
\end{equation}
so that ${\rm det}[V(-k_x)]={\rm det}[V(k_x)]$.

\paragraph*{Removing the U(1) gauge discontinuity.}

Obviously, $V(k_x)\in \mathrm{U}(2)$, which can always be written
as a $\mathrm{U}(1)$ phase times an $\mathrm{SU}(2)$ matrix.
For our next step, we find it convenient to reduce $V(k_x)$ to
$\mathrm{SU}(2)$ form by multiplying the states $|u_{n\k}\rangle$
by a $\k$-dependent phase factor.  To do so, we define
\begin{equation}
\gamma(k_x)=\mathrm{Im}\log \det V(k_x)
\end{equation}
with the branch choice that $\gamma=0$ at $k_x$=0 and
$\gamma(k_x)$ is a continuous function of increasing $k_x$.
This results in $\gamma=0$ again at $k_x=2\pi$ because
the TR symmetry forces the total Chern number
$C$ of the two bands to be zero.  Indeed, $C$ is just given
by the winding number of the $\mathrm{U}(1)\rightarrow\mathrm{U}(1)$
mapping from $k_x$ to $\gamma$.  This follows from
\begin{eqnarray}
2\pi C &=& \int_0^{2\pi}dk_x \, \left[
   \mathrm{Tr}\,{\cal A}_{k_x}^{(k_y=-\pi)}-
   \mathrm{Tr}\,{\cal A}_{k_x}^{(k_y=\pi)} \right] \nonumber\\
&=&   \int_0^{2\pi}dk_x \, \mathrm{Im}\mathrm{Tr}
   \left[ V^\dagger \partial_{k_x} V\right] \nonumber\\
&=& \int_0^{2\pi}dk_x \, \partial_{k_x} \gamma(k_x) \nonumber\\
&=&\gamma(k_x) \Big\vert^{2\pi}_0
\label{cproof}
\end{eqnarray}
after some algebra.

Thus, our next step is simply to shift the phases of all
states according to
\begin{equation}
|u_{n(k_x,k_y)}^{\rm new}\rangle = e^{-i\gamma(k_x)k_y/4\pi}
|u_{n(k_x,k_y)}\rangle .
\end{equation}
This conserves all of the previous properties (smooth gauge inside
the cylindrical BZ and on all boundaries except at $k_y=\pm\pi$).
Moreover, $V(k_x=0)$ is still the identity, but now in addition,
$\mathrm{det}\,V(k_x)$ is real and positive at all $k_x$.  That is,
$V(k_x)$ has been reduced to $\mathrm{SU}(2)$ form.
We also note that Eqs.~(\ref{chi0}) and (\ref{VV})
continue to hold.
However, $V(k_x)$ remains off-diagonal at general $k_x$,
thus signaling that the decomposition of the occupied
subspace into the direct sum of the two TR-symmetric
subspaces is not yet complete.

As noted earlier, the fact that our procedure starts from
Kramers-degenerate pairs at $(k_x,k_y)=(0,0)$ and $(\pi,0)$ and
respects TR symmetry at all stages enforces that $V(k_x)$
must be a constant times the identity at $k_x=0$ and $k_x=\pi$.
Since $V\in\mathrm{SU}(2)$ as well, $V$ must be $I$ or $-I$
at these two $k_x$ values.  Previous gauge-fixing choices insure
that $V(0)=I$, but is $V(\pi)=I$ or $-I$?  It can be shown that
these choices correspond to the case of the $Z_2$ index being even
or odd, respectively.  Indeed, according to homotopy theory, the
mapping $\mathrm{U(1)} \rightarrow\mathrm{SU(2)}$ is characterized
by a $Z_2$ index; this is precisely the case here.  In fact, the
procedure up to this point can be used as an alternative to the
method we presented earlier in Ref.~\onlinecite{Soluyanov-PRB11-b}
to compute the $Z_2$ invariant.
From the numerical perspective, however, such a method does
not have any significant advantages compared to the previously
suggested one, apart from its straightforward geometric
interpretation. In fact, for large systems it might not be very
convenient to carry out all the transformations of the
wavefunctions described above.

In what follows, we assume that the $Z_2$ index is odd.

\subsection{Disentangling the two bands}

In order to proceed, we want to make $V(k_x)$ diagonal at each $k_x$.
When this is accomplished we will have two disentangled bands
1 and 2, although each will still have its own phase
discontinuity along the boundaries at $k_y=\pm\pi$.  We take
a first step in this direction by taking advantage of the
freedom that we had when choosing the initial representatives of
the occupied subspace at ${\bf k}=(0,0)$. These two states may
be changed by a unitary transformation ${\cal U}$,
which we take to belong to $\mathrm{SU(2)}$ so that the TR
symmetry is fully preserved. So, we first look for the global
$\mathrm{SU(2)}$ rotation that will minimize the sum of all the
off-diagonal terms of the $V$ matrices at all $k_x$. Once this
is done, a further adjustment can be made so as to make $V(k_x)$
exactly diagonal at each $k_x$ without losing
smoothness on the cylinder. We now explain the procedure
in detail.

\subsubsection{Steepest-descent minimization of ${\cal V}_{\rm OD}$}

Let us introduce a functional
\begin{equation}
{\cal V}_{\rm OD}=\frac{1}{N_x}\sum_{k_x} \sum_{m\neq n} |V_{mn}(k_x)|^2
\label{vod}
\end{equation}
that is a measure of the degree to which $V(k_x)$ fails to be diagonal
along the discontinuity at $k_y=\pm\pi$.  The sum on $k_x$ runs over
a uniform grid of $N_x$ mesh points.
%
%\dvm{Reworded.}
%
We want to use the freedom of choosing the initial pair of states
at ${\k}=(0,0)$ to minimize this functional by rotating the states
at all $k$-points by the same unitary matrix ${\cal U}_0$.  To do
so, we consider the gradient of ${\cal V}_{\rm OD}$ with respect to
an infinitesimal $k$-independent unitary transformation
\begin{equation}
U_{mn}=\delta_{mn}+dW_{mn},
\label{inf}
\end{equation}
where $dW=-dW^\dagger$ for $U$ to be unitary.
A transformation of this form rotates the states according to
\begin{equation}
|\tilde{u}_{n{\bf k}}\rangle=|u_{n{\bf k}}\rangle
  +\sum_m dW_{mn} |u_{m{\bf k}}\rangle.
\end{equation}
To first order in $dW$ the change in $V(k_x)$ is
%
%\dvm{Wrote commutator instead.  OK?}
%
\begin{equation}
dV_{mn}=\left[V,dW\right]_{mn} \,.
\label{dv}
\end{equation}
To compute the gradient
%
%\dvm{Changed according to your suggestion.}
%
\begin{equation}
G_{mn}=\left(\frac{d{\cal V}_{\rm OD}}{dW}\right)_{mn}
=\;\frac{d{\cal V}_{\rm OD}}{dW_{nm}}
\label{gradient}
\end{equation}
we note that Eq.~(\ref{vod}) can be rewritten in the form
\begin{equation}
{\cal V}_{\rm OD}={\cal N}-\frac{1}{N_x}\sum_{k_x}
\sum_n^{\cal N} |V_{nn}(k_x)|^2.
\label{newvod}
\end{equation}
Then, using  Eq.~(\ref{dv}), one can write
\begin{eqnarray}
d{\cal V}_{\rm OD}&=&-\frac{2}{N_x}\mathrm{Re}
\sum_{k_x} \sum_{nm} V_{nn}^*(V_{nm}dW_{mn}- dW_{nm} V_{mn})
\nonumber\\
&=& -\frac{2}{N_x}\sum_{k_x} \mathrm{Re\,Tr}\left[R(k_x)\,dW\right]
\label{dV}
\end{eqnarray}
(the $k_x$ dependence of $V$ is suppressed for brevity) and
\begin{equation}
R_{mn}(k_x)=V_{nm}[V_{nn}^*-V_{mm}^*] .
\end{equation}
The second line of Eq.~(\ref{dV}) is obtained by
interchanging the dummy $nm$ indices in the second term of the
first line. It then follows that
%
%\dvm{I tried to shorten this whole discussion.  The reader
%should get the main idea without getting too stuck in details.}
%
\begin{equation}
G=\frac{1}{N_x}\sum_{k_x}\left[R(k_x)-R^\dagger(k_x) \right].
\end{equation}

We emphasize that the gradient $G$ is independent of $k_x$ since
it generates a global unitary rotation to be applied simultaneously
to all states.  Also, $G$ is not only antihermitian but also
traceless, so that it generates a $\mathrm{SU(2)}$ unitary
rotation.  We now follow an iterative steepest-descent procedure,
choosing a small positive damping constant $\beta$ and letting
%
%\dvm{You are correct; I fixed this part a little, putting back one
%of your equations that I removed before (but inline now).  Is it
%OK?}
%
$dW=-\beta G^\dagger$ (i.e, $dW=\beta G$) so that
$d{\cal V}_{\rm OD}=\mathrm{Tr}[G\,dW]=-\beta||G||^2$ to first
order in $\beta$.
We use this to update the states according to
\begin{equation}
|u_n^{(j+1)}\rangle = \sum_m \left[e^{\Delta W^{(j+1)}}
\right]_{mn}|u_n^{(j)}\rangle
\end{equation}
and the $V$ matrices according to
\begin{equation}
V^{(j+1)}=\left[e^{\Delta W^{(j+1)}}\right]^\dagger
V^{(j)} e^{\Delta W^{(j+1)}}
\end{equation}
where the upper index refers to the iteration
step.  The iteration stops when
${\cal V}_{\rm OD}^{(j)}-{\cal V}_{\rm OD}^{(j+1)}$
stays consistently below some pre-chosen tolerance $\varepsilon$.

To give a flavor of how steepest descent works we give
the values obtained for the Kane-Mele model in the QSH regime
($\lambda_v/t=1$, $\lambda_{SO}/t=0.6$, $\lambda_{R}/t=0.5$)
with a $120\times120$
$k$-mesh, $\varepsilon=10^{-6}$ and $\beta=0.25$.
Initially ${\cal V}_{\rm OD}=0.0226$,
while after minimization
${\cal V}_{\rm OD}=0.0021$, so it becomes approximately
ten times smaller. The crucial thing is that this
final value of ${\cal V}_{\rm OD}$ suggests that the average
off-diagonal element of $V$ has is of order $\times10^{-2}$,
meaning that the $V$ matrix is almost diagonal.

\begin{figure}
\begin{center}
\includegraphics[width=2.8in]{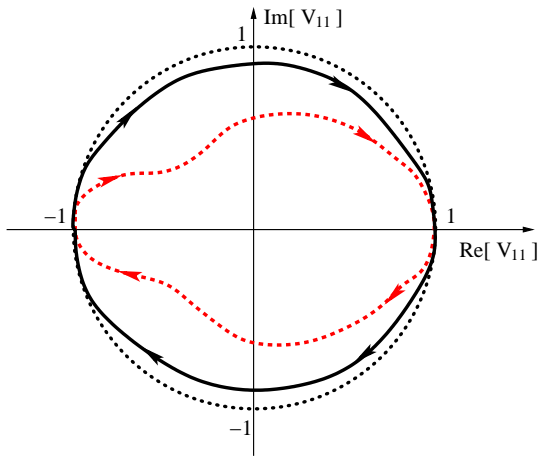}
\end{center}
\caption{Trajectory of $V_{11}$ in the complex plane as $k_x$
runs across the BZ, before (red dashed line) and after (solid
black line) the global
${\cal U}_0$ rotation that minimizes ${\cal V}_{\rm OD}$
for a $\z2$-odd insulator. In neither case is
the graph exactly a unit circle (dotted line),
but $V_{11}(0)=1$ and $V_{11}(\pi)=-1$.}
\label{fig:3}
\end{figure}

Note that at this stage the two subspaces are still
not completely disentangled into two well-defined
Chern subspaces. However, the gauge is very close to
what we need. For example, the winding of $V(k_x)$ already
has the necessary features: if one plots $V_{11}$ in the
complex plane as a function of $k_x$, one will see that it
winds once around the origin in the counterclockwise
direction as $k_x$ goes from $0$ to $2\pi$, as illustrated
in Fig.~\ref{fig:3}.
Since $V$
are not diagonal yet the trace is not the unit circle,
although it is close.  $V_{22}$ winds in the opposite direction.

\subsubsection{Diagonalization and final decomposition}
Now we are in a position to make the final step
in decomposition procedure. As a result of the steps above,
the off-diagonal elements of the
$V(k_x)$ matrices should be small compared to the
diagonal ones, so that the matrices are almost diagonal.
This means that $V(k_x)$ can be diagonalized by a unitary
transformation ${\cal U}(k_x)$ that is only slightly different
from the unit matrix.  Since diagonalization of $V(k_x)$ does
not fix the phases of the eigenvectors, and we need the phases
to vary smoothly, we need an extra step to fix these phases.
We do this by enforcing that the dominant component
of each eigenvector of $V(k_x)$ is real and positive.\footnote{
  As an alternative, one could carry out a single-band parallel
  transport of the two resultant states along $k_x$ to smooth out
  the random phase variations at different $k_x$ introduced by
  the diagonalization procedure.}
%
%\dvm{Fine.}
%
We then apply ${\cal U}(k_x)$ to rotate the states at
all $k_y$ for each given $k_x$ (except at $k_x=0$ or $\pi$,
where $V$ was already diagonal).

%\dvm{Tried to condense around here.}
%
As a result of this step the occupied subspace has been disentangled
into a direct sum of two subspaces corresponding to states
$n=1$ and 2.  Moreover, they should form Kramers pairs and
satisfy the constraint~(\ref{chi0}).
Each subspace has a gauge that is smooth on the cylinder but not
on the torus, since there is still a phase mismatch, corresponding
to $V_{nn}(k_x)$, across the boundary at $k_y=\pm\pi$.  For the
$Z_2$-odd case this phase discontinuity can never be completely
removed, since the subspaces have Chern numbers of $\pm1$.

To check the procedure, we apply it to the Kane-Mele model
and compute the individual Chern numbers of the two disentangled
bands. The computation is done for each band separately
using the Abelian definition of Berry curvature, Eq.~(\ref{C1def}).
The result is $C_1=-1$ and $C_2=+1$. The fact that the two
states have well-defined Chern numbers is a signature of
disentanglement, so that the individual Chern numbers of
Eq.~(\ref{indChern}) have integer values ($c_1=-c_2=-1$).
The TR constraint of Eq.~(\ref{chi0}) is indeed respected
at each $k$-point. Thus we conclude that we have succeeded in
finding a decomposition of the occupied subspace into a direct sum
of two Chern subspaces that are mapped onto each other by the
TR symmetry. Once again, we see that the TR-symmetric
gauge for topological insulators is discontinuous on the BZ
torus.

\subsection{Establishing a cylindrical gauge}

%\dvm{Tried to write better introductory paragraph.}
%
In Sec.~\ref{sec:single} we introduced a special ``cylindrical
gauge'' for which the states satisfy Eq.~(\ref{1band2D}).
The defining characteristic of this special gauge is that
the phase discontinuity at the cylinder boundary evolves at a
\textit{constant rate} as a function of $k_x$.  As we shall see
in Sec.~\ref{sec:rotation}, it is useful to have such a ``standard
gauge'' enforced on the states when using them in some subsequent
operations.  Here we show how to extend our procedure so as to
conform to the requirements of the cylindrical gauge.

As was mentioned above, the diagonal elements of
$V(k_x)$ wind around zero
in the complex plane in opposite directions, changing by
$2\pi$ when $k_x$ goes from $0$ to $2\pi$. Since we have carried out
the diagonalization of the $V$ matrices, we know that the $V_{jj}$
elements follow a unit circle in the complex plane
of the form $e^{i\rho_j(k_x)}$. However, the speed of this rotation
given by $v_j(k_x)=d\rho_j/dk_x$ (where $\rho_j$ remains on the
same branch of the logarithm) is not constant, in contrast to the
requirement of the cylindrical gauge.

To change the speed of winding of $V(k_x)$
we apply the gauge transformation
\begin{equation}
W(k_x,k_y)=\left[V_{\rm targ}(k_x) V^\dagger(k_x)\right]^{k_y/2\pi}
\end{equation}
to the the occupied states at each $(k_x,k_y)$. Here
\[
V_{\rm targ}(k_x)= \begin{pmatrix}
e^{ik_x c_1} & 0  \\
0 & e^{ik_x c_2}  \end{pmatrix}.
\]
gives the target shape of $v$ that corresponds to the
cylindrical gauge.
Note that the choice of sign should be correlated with the
individual Chern number of the band it is applied to.
Such a gauge transformation is obviously continuous on the cylinder
and does not change the topology of the individual bands.
It also preserves the TR symmetry of the states and the
relation of Eq.~(\ref{chi0}) is still satisfied.

We note that if the above decomposition
is applied to a normal insulator (say, the Kane-Mele model
in the normal-insulator regime), then $c_1=c_2=0$ and
a smooth gauge is obtained at this step.

\subsection{Relation to spin Chern numbers}

Finally, we would like to compare our approach to
disentangling $\z2$ bands into
Chern bands to some other approaches suggested previously.
In the work of Ref.~\onlinecite{Sheng-PRL06} the authors
suggested to associate a Chern number with each
possible spin projection value. This is especially
convenient when $\hat{s}_z$ is conserved; then
it is natural to assign individual Chern numbers to
each of the bands identified by a particular value of $s_z$. Such
Chern numbers were called ``spin Chern numbers.'' For
example, in the case of the Kane-Mele model with no Rashba
coupling (i.e., $\lambda_R=0$), $\hat{s}_z$ is conserved and
the Hamiltonian becomes block-diagonal with respect
to the spin projection, allowing for well-defined spin Chern
numbers. When the Rashba interaction is turned on
the mirror symmetry of the model is broken and $\hat{s}_z$
is no longer conserved, thus making the original concept
of a spin Chern number obscure.

This issue was clarified further by
Prodan,\cite{Prodan-PRB09} who showed
that even with the spin-mixing Rashba term it is possible to
define spin Chern numbers by diagonalizing $\hat{s}_z$ in the
occupied space of $\z2$ insulator at each $\k$. In other words,
one diagonalizes the operator $\hat{P}_{\k}\hat{s}_z\hat{P}_{\k}$,
where $\hat{P}_{\k}$ is the projector onto the occupied states
at $\k$. Then, if the eigenvalues
turn out to be separated by a spectral gap from one another at
each value of $\k$, one can identify these ``bands'' as the desired
manifolds, and carry out a unitary rotation of the original bands
into these states to disentangle them. The spin Chern numbers
thus defined for these bands
are well defined and, in fact, correspond to the individual
Chern numbers of our work. However, when the spectral gap
between any two eigenvalues of the projected spin operator
closes, such a decomposition becomes impossible. One could
still consider some other projection operators based on mirror
or other symmetries, as in Ref.~\onlinecite{Teo-PRB08},
and use these eigenvalues in a similar way to disentangle
the occupied states.  However,
such a method always relies on some symmetry of a particular model,
and is thus not universal. The method suggested in the present work,
in contrast,
does not depend on any symmetries of the underlying
system. Thus, we conclude that individual Chern numbers
proposed in the present work are robust and arise solely
from the topology of the occupied subspace of the system.

Finally, it was discussed elsewhere that the spin Chern numbers
do not contain any more information than the $\z2$ invariant,
because their sign can be changed without closing the
insulating gap.\cite{Fu-PRB06, Fukui-PRB07, Prodan-PRB09}
This is the case for individual Chern numbers as well,
since obviously, one can simply change the labeling
of the states by a simple unitary transformation
that interchanges $|u_{1\k}\rangle$ with $|u_{2\k}\rangle$.
Therefore, individual Chern numbers are merely an alternative
way of describing the occupied subspace of a
$\z2$ insulator in terms of disentangled
bands, and do not contain any more information about
the topological state of the whole system than a $\z2$
invariant alone.

\section{Rotation into a smooth gauge}
\label{sec:rotation}
We now discuss the final step in our construction of
a smooth gauge for a QSH insulator starting from
the two Chern bands obtained at the previous steps.
The task of unwinding the topological twists of these
bands requires a unitary transformation
that is also topologically nontrivial in the following
sense.
Obviously, a transformation that is smooth on the BZ torus,
being periodic in the $k_y$ direction, cannot make a cylindrical
gauge smooth. One needs instead a unitary
transformation ${\cal G}(\k)\in{\mathrm U(2)}$ that has a discontinuity
on the torus that exactly cancels out the discontinuities
of the cylindrical-gauge states.
%
%\dvm{Notation of $\cal G$ introduced above.  Also, sentence added
%assuming next paragraph can be dropped.}
%
Of course, since the total Chern number of the whole
occupied space is a topological invariant,\cite{Avron-PRL83}
the transformation
will preserve the condition that the total Chern number is zero.
In particular, the rotation we are looking
for makes $c_1=c_2=0$.

A unitary transformation that solves the problem of unwinding
the two QSH bands with Chern numbers $\pm 1$ is given naturally by the
solution of the Haldane model\cite{Haldane-PRL88} of a Chern
insulator (CI), or for that matter, of any two-band model of a CI.
Indeed, the unitary transformation ${\cal G}({\bf k})$ that
diagonalizes the Hamiltonian in that case is one that rotates the
two topologically trivial tight-binding basis states
$(1,0)^T$ and $(0,1)^T$ into the
eigenstates of the model.
Obviously, ${\cal G}^{-1}({\bf k})={\cal G}^\dagger({\bf k})$
rotates the topologically nontrivial states back into the
trivial ones, and thus can be used to unwind our QSH states.
%
%\dvm{I prefer to briefly mention the cylindrical gauge condition first,
%before going into details.}
%
In order for this procedure to produce a smooth gauge, the
Hamiltonian eigenstates of the CI model also have to be smoothly
defined on the cylinder and obey the same cylindrical gauge of
Eqs.~(\ref{1band2Dpsi}-\ref{1band2D}).  Assuming this has been
done, the application of the resulting ${\cal G}^\dagger(\k)$
to the QSH states defined by our procedure will finally result in
a gauge that is smooth everywhere on the torus and that generates
new bands with $c_1=c_2=0$, as desired.

The numerical implementation of this procedure is done
most conveniently by solving the CI model on the same 2D $\k$-space
mesh as was used to solve for the QSH states.  If the latter
have been computed in the context of first-principles calculations
or of some complex tight-binding model, then some
known CI model such as the Haldane model can be used to
provide the needed ${\cal G}(\k)$.  However, when working with
a minimal $4\times4$ tight-binding model for a QSH system,
it may be more convenient to use a
$2\times 2$ spin-up (or spin-down) block of the
original $4\times 4$ QSH model itself.  After all,
this already lives on the needed $\k$-mesh and generates bands
with Chern numbers of $\pm 1$.
For example, for an application to the Kane-Mele model
in the QSH regime ($\lambda_v/t=1$,
$\lambda_{SO}/t=0.6$, $\lambda_{R}/t=0.5$),
we used the spin-up block of the original
Hamiltonian and obtained two states $|{u}_{i\k}^\prime\rangle$
with Chern numbers ${c}_1^\prime=-1$ and ${c}_2^\prime=1$, where the
hat is used to distinguish the CI quantities from the QSH ones.

As mentioned earlier, it is also necessary to bring the CI bands
$|{u}^\prime_{i\k}\rangle$ into the cylindrical gauge in order to
ensure that the resulting ${\cal G}^\dagger(\k)$ exactly cancels
the discontinuity of the QSH bands at the edge of the
cylinder.  For this purpose, a parallel-transport procedure
is carried out across the BZ in close analogy to what was
described in Sec.~\ref{subs:movg}, but now it is done in a single-band
$\mathrm{U(1)}$ context applied to each of the CI states in turn.
It is useful to refer again to Fig.~\ref{fig:2}.  First, a parallel
transport of $|{u}^\prime_{1\k}\rangle$ is carried out along the $k_x$
axis (with an arbitrary choice of phase at $\k=0$), and a graded
phase twist is applied to match phases at $k_x=0$ and $2\pi$ as
in Figs.~\ref{fig:2}(b-c).  Then parallel transport is performed
along the vertical directions as in Fig.~\ref{fig:2}(d), and a
($k_x$-independent) phase change that is graded along $k_y$ is
applied to restore continuity at the corner points of Fig.~\ref{fig:2}(e).
This defines a phase discontinuity ${V}^\prime_{11}(k_x)= \langle
{u}^\prime_{1}(k_x,-\pi)| {u}^\prime_{1}(k_x,\pi)\rangle$ whose phase-winding
rate $d\,\ln(V_{11})/dk_x$ is initially nonuniform, but is
made uniform by the same trick as for the QSH states.  The
procedure is repeated for the second CI band.

The above procedure results in Chern bands obeying the cylindrical
gauge as required.  We can now simply form
the desired unitary matrix ${\cal G}(\k)$ as the
$2\times2$ matrix whose first and second columns are filled
with the column vectors $|{u}^\prime_1(\k)\rangle$ and
$|{u}^\prime_2(\k)\rangle$ respectively.  We emphasize again that
this matrix is not topologically trivial; its
coefficients are continuous on the cylinder,
but not continuous across $k_y=\pm\pi$, just like
the CI that has produced it.
Applying ${\cal G}^\dagger({\bf k})$
to the QSH bands constructed in Sec.\ref{sec:decomp},
\begin{equation}
|\tilde{u}_{n\k}\rangle=\sum_m {\cal G}^\dagger_{mn}(\k)|u_{m\k}\rangle \,,
\end{equation}
we finally end up with two bands
that have $c_1=c_2=0$ and that span the Hilbert space defined by
the original occupied bands of the QSH model.  Thus, we have
constructed a smooth and periodic gauge for the target $\z2$
insulator.

It should be stressed that rotation into a smooth
gauge as described above breaks TR symmetry,
since ${\cal G}(\k)$ results from a TR-broken CI model.
Thus, the two smooth subspaces are not mapped onto each other
by the TR operator, so that $\langle \tilde{u}_{1,\k}
|\theta|\tilde{u}_{2,-\k}\rangle \ne 0$ except at
TR-invariant momenta $\k=-\k+\G$.
Similarly, if Wannier functions are constructed from the
Bloch spaces defined in this way, they will not form
Kramers pairs.\cite{Soluyanov-PRB11-a}
Finally, we note that although the gauge is now smooth and periodic,
it can be smoothed further by using this gauge as a starting
point for a Wannier-function maximal-localization
procedure.\cite{Marzari-PRB97}

In summary, we have demonstrated a general method for constructing
a smooth gauge for a $\z2$ topological insulator.  At this
final stage we start with a gauge that still respects TR
symmetry, but then we carry out a unitary mixing operation that
violates this symmetry in order to avoid the topological
obstruction.  Application to the Kane-Mele model allows us to
compute the $\z2$ invariant with the smooth-gauge formula of Fu
and Kane\cite{Fu-PRB06} as discussed in Appendix~\ref{app:trs}.

\section{Conclusions}
\label{sec:concl}
In this paper we have developed a general method for
decomposing the occupied space of a $\z2$ insulator
into a direct sum of two TR-symmetric Chern subspaces
with nontrivial individual Chern numbers. We then
described a general procedure for breaking the TR symmetry
between the two bands and rotating them into subspaces that
are smooth everywhere on the torus.  Our methods are
general in the sense that they do not make use of any special
symmetries or assumptions about gaps in the spectrum of spin
operators.  This establishes the construction of a smooth gauge
for 2D topological insulators.

%\dvm{\scr ``We think that in 3D such a procedure might become too complicated
%and a different approach should be developed. In particular,
%an approach that would connect the decomposition proposed
%in the present paper with a well established concept of
%spreads of the Wannier functions\cite{Marzari-PRB97}, might
%well work.''\\
%AS: Actually, isn't generalization to 3D trivial in some sense?
%Imagine a 3D cubic BZ of a strong (weak) $\z2$ insulator. We just
%apply our procedure to the 2D surfaces that correspond to an odd
%2D $\z2$ index, and then connect all the surfaces somehow.
%I guess, even parallel transport in the third direction might work,
%if the topologically
%nontrivial surfaces are chosen correctly by some particular
%choice of reciprocal vectors.\\
%DV: I'm not sure I agree.  In any case, it's a tricky issue.
%if we want to discuss it, it should probably not be the final
%paragraph of the Conclusions.  I think my inclination is not
%to mention it.}
%

\begin{acknowledgements}
We would like to thank C. L. Kane and
E. Prodan for useful discussions.
This work was supported by NSF Grant DMR-1005838.
\end{acknowledgements}

\appendix

\section{Parallel transport}
\label{app:partr}

Let us discuss how to construct a parallel-transport gauge starting
from a set of randomly chosen eigenstates of the Hamiltonian on a
$\k$-mesh. In what follows we distinguish single-band
and multiband parallel transport procedures. The general idea
in both cases is to carry the Bloch states along a certain path
in the BZ in such a way that they remain as parallel as possible
to the previous states at all points. If the path is closed,
the states
might return to the initial point with some phase differences
relative to the initial states, thus violating singlevaluedness.
However, singlevaluedness of the wavefunction can be restored by
spreading the extra phase uniformly along the path, as
explained in more detail below. For simplicity, we consider
parallel transport along one direction in the BZ, say $k_x$.
In this case, a closed loop is obtained when the state is
transported by a reciprocal lattice vector $\G_x$.
The generalization to an arbitrary direction should be obvious.

Consider a single isolated band $|u_{n\k}\rangle$. To carry the
state to $k+\Delta k$ via parallel transport, the phase of the
Bloch state at this new point should be chosen in such a way that
the overlap $\langle u_{n\k}| u_{n,\k+\Delta k_x}\rangle$ is real
and positive, so that the
change in the state is orthogonal to the state itself.
It is straightforward to implement this numerically. Consider a discrete
uniform mesh of $k$-points $\{\k_j\}, j\in[1,N+1]$, where
$\k_{j+1}=\k_j+\Delta k_x$ and $\k_{N+1}=\k_1+\G_x$.
The states $|\tilde{u}_{\k_j}\rangle$ at these points are obtained by
a numerical diagonalization procedure and thus have random phases.
At the initial point $j$=1 we set
$|u'_{\k_1}\rangle =|\tilde{u}_{\k_1}\rangle$.  Then at each
subsequent $\k_{j+1}$ we let $\beta_{j+1}={\rm Im}\,\ln\,
\langle \tilde{u}_{\k_{j+1}}^{\phantom{,}}|\, u'_{\k_j} \rangle$ and
then apply the ${\cal U}(1)$ phase rotation
\begin{equation}
|u'_{\k_{j+1}}\rangle=e^{i\beta_{j+1}}|\tilde{u}_{\k_{j+1}}\rangle \,,
\end{equation}
which makes $\langle u'_{\k_j}|u'_{\k_{j+1}}\rangle$ real and
positive.  Once this is done at each $\k$-point,
the state at $\k_1$ differs from that at $\k_{N+1}$ by a phase factor
$e^{i\phi}$, where $\phi$ is chosen on a particular branch, say
$\phi\in(-\pi,\pi]$. $\phi$ is the Berry phase associated
with the traversed path.
Unless $\phi=0$, periodicity in $k_x$ is lost.
To restore it, the extra phase should be spread uniformly along
the string of $\k$-points, i.e.,
\begin{equation}
|u_{\k_j}\rangle=e^{-i\phi \k_j/2\pi}|u'_{\k_j}\rangle
=e^{-i(j-1)\phi/N} |u'_{\k_j}\rangle,
\label{t1}
\end{equation}
where in the last equality the uniformity of the $k$-mesh was used.

In the multiband case one deals with the non-Abelian generalization of
the Abelian Berry phase.\cite{Wilczek-PRL84, Mead-RMP92}
We now consider an isolated set of ${\cal N}$ bands and describe
parallel transport in the $k_x$-direction in the non-Abelian
case.\cite{Marzari-PRB97, Resta-JPC00} The parallel transport
gauge is constructed by requiring that the overlap matrix
\begin{equation}
\tilde{M}_{mn}^{(\k_j,\k_{j+1})}=\langle \tilde{u}_{m{\k_j}}| 
\tilde{u}_{n \k_{j+1}}\rangle
\end{equation}
must be Hermitian, with all positive eigenvalues, at each step.
This is uniquely accomplished by means of the singular value
decomposition in which an ${\cal N}\times{\cal N}$ matrix $M$
is written in the form $M=V\Sigma W^\dagger$, where $V$ and $W$
are unitary and $\Sigma$ is positive real diagonal. If the
states at $\k_{j+1}$ are rotated by ${\cal U}=WV^\dagger$, i.e.,
\begin{equation}
|{u}^\prime_{n\k_{j+1}}\rangle=\sum_m^{\cal N}
{\cal U}_{mn}(\k_{j+1})|\tilde{u}_{m\k_{j+1}}\rangle,
\end{equation}
the new overlap matrix ${M}_{mn}^{\prime\, {(\k_j,\k_{j+1})}}$
will be of the form $V\Sigma V^\dagger$, which is Hermitian with
positive eigenvalues as desired.
Repeating this procedure up to $j=N$, one obtains that the new
states $|u^\prime_{n\k_{N+1}}\rangle$ are related to the states
$|u^\prime_{n\k_1}\rangle$ by a unitary transformation $\Lambda$
according to
\begin{equation}
|u^\prime_{n{\k_1}}\rangle=
e^{2\pi ix}
\sum_m^{\cal N}\Lambda_{mn}|u^\prime_{m{\k_{N+1}}}\rangle.
\label{lambda}
\end{equation}
The eigenvalues of this matrix are of the form
$\lambda_n=e^{-i\phi_n}$, where the phases $\phi_n={\rm
Im\,ln}\,\lambda_n$ (again chosen according to some definite branch
cut) are the analogs of the Abelian Berry phases.

To restore periodicity we follow the same trick as
in the single-band case, but generalized to the matrix form.
To do this one finds the unitary matrix $R$ that diagonalizes
$\Lambda$, and then rotates all states at all $\k_j$ by this
same unitary $R$, so that the new states correspond to a diagonal
$\Lambda$ with its eigenvalues $\lambda_n=e^{i\phi_n}$ on the
diagonal.  Now it is straightforward to
obtain periodicity by applying the graded phase twists
\begin{equation}
|{u}_{n\k_j}\rangle= e^{-i(j-1)\phi_j/N}|u^\prime_{n\k_j}\rangle.
\label{t2}
\end{equation}
This results in a gauge that is smooth along $k_x$ and $\G_x$-periodic.

\section{Kane-Mele model}
\label{app:KMmodel}

Here we briefly summarize the Kane-Mele model~\cite{Kane-PRL05-b}
of a quantum spin Hall system. This model is represented
by a tight-binding (TB) Hamiltonian on a honeycomb lattice with
dimensionless lattice vectors
$a_{1,2}=(\sqrt{3}\hat{\bf y}\pm\hat{\bf x})/2$.
The Hamiltonian is
$$
H=\lambda_v\sum_i \xi_i c_i^\dagger c_i +
\sum_{<ij>} c_i^\dagger \left(t+i\lambda_R[{\bf s}
          \times{\hat {\bf d}}_{ij}]_z\right)c_j
$$
\begin{equation}
+i\lambda_{SO}\sum_{\ll ij \gg}\nu_{ij} c_i^\dagger s^z c_j,
\label{hamKM}
\end{equation}
where the three terms represent on-site, first-, and second-neighbor
interactions respectively. Here $\xi_i=\pm 1$
represents a staggered on-site interaction (breaking the inversion
symmetry of the original honeycomb lattice), and $\lambda_{SO}$ and
$\lambda_R$ represent the effects of spin-orbit interaction
(the latter breaks $S_z$ conservation and violates mirror symmetry
in the $xy$-plane). Also, $\hat{\bf d}_{ij}$ is a unit vector
directed from site $i$ to site $j$, while
$\nu_{ij}=(2/\sqrt{3})[\hat{\bf d}_1\times \hat{\bf d}_2]=\pm 1$, where
$\hat{\bf d}_1$ and $\hat{\bf d}_2$ represent the directions of the
two bonds along which the electron hops in going from site $i$ to
site $j$.

Using the TB convention
$\chi_{j\sigma\k}(\r)=\sum_{\R}e^{i\k\cdot\R}\varphi_{s}(\r-\R-
{\bf t}_{j})$, where $\varphi$ are TB basis functions, $s$ stands
for the spin index, and ${\bf t}_j$ is the vector from the origin to
the $j$-th atom in the home unit cell, the Hamiltonian is
written as
\begin{equation}
H({\bf k})=\sum_{\alpha=1}^5 d_{\alpha} ({\bf k}) \Gamma^{\alpha} +
\sum_{\alpha<\beta=1}^5 d_{\alpha \beta} ({\bf k}) \Gamma^{\alpha \beta}.
\label{Dirac-Hamiltonian}
\end{equation}
Here the Dirac matrices are
$\Gamma^{1,2,3,4,5} = (I\otimes \sigma^x, I\otimes \sigma^z, s^x
\otimes \sigma^y, s^y \otimes \sigma^y, s^z \otimes \sigma^y)$
with the Pauli matrices $\sigma^k$ and $s^k$ acting in sublattice
and spin space respectively, and the commutators are $\Gamma^{\alpha
\beta}=[\Gamma^\alpha, \Gamma^\beta]/(2i)$. The original
reciprocal-lattice coordinates ${\kappa_1}$ and $\kappa_2$ may be changed
into $k_x\in[0,2\pi]$ and $k_y\in[-\pi,\pi]$ via
$k_x = \kappa_x/2-\sqrt{3}\kappa_y/2$ and $k_y =
\kappa_x/2+\sqrt{3}\kappa_y/2$.  The resulting $d$ coefficients
are given in Table~\ref{Tab:coefficients}.
\begin{table}[ht]
\begin{center}
\begin{tabular}{c c c c}
\hline
\hline
$d_1$ & $t(1+2\cos{\alpha}\cos{\beta})$               &
$d_{12}$ & $ -2t\cos{\alpha}\sin{\beta}$                      \\
$d_2$ & $\lambda_v$                          &
$d_{15}$ & $2\lambda_{SO}(\sin{2\alpha}-2\sin{\alpha}\cos{\beta})$ \\
$d_3$ & $\lambda_R(1-\cos{\alpha}\cos{\beta})$        &
$d_{23}$ & $-\lambda_R\cos{\alpha}\sin{\beta}$                 \\
$d_4$ & $-\sqrt{3}\lambda_R \sin{\alpha}\sin{\beta}$ &
$d_{24}$ & $\sqrt{3}\lambda_R \sin{\alpha}\cos{\beta}$         \\
\hline
\hline
\end{tabular}
\end{center}
\caption{Nonzero coefficients in Eq.~(\ref{Dirac-Hamiltonian}).
Here $\alpha=(k_x+k_y)/2$ and
$\beta=(k_y-k_x)/2$ with  $k_x = \kappa_x/2-\sqrt{3}\kappa_y/2$
and $k_y = \kappa_x/2+\sqrt{3}\kappa_y/2$. The lattice constant
is assumed to be of unit length. }
\label{Tab:coefficients}
\end{table}
This model respects time-reversal symmetry and realizes the QSH
regime, i.e., it represents a 2D $\z2$ topological insulator in
some regions of its parameter space.\cite{Kane-PRL05-b}
For our illustrative tests we have used
$\lambda_v/t=1$, $\lambda_{SO}/t=0.6$ and $\lambda_{R}/t=0.5$ for
the topological phase, and have changed
$\lambda_v/t$ to $5$ to access the normal phase.

\section{Time-reversal constraint and smooth gauge}
\label{app:trs}

In Ref.~\onlinecite{Fu-PRB06} Fu and Kane developed a theory of
a $\z2$ periodic spin pump of a 1D insulating
system.
That work established a formula for computing
the $\z2$ invariant given a smooth gauge.
In this Appendix we review this result and discuss it
from the perspective of the smooth gauge constructed in
the present work.

The work of Ref.~\onlinecite{Fu-PRB06} focuses on the
pumping process in 1D gapped periodic Hamiltonians subject to
the conditions $H(t+T)=H(t)$ and $H(-t)=\theta H(t)\theta^{-1}$,
where $t$ is the pumping parameter. Such a pump becomes TR-invariant
at $t=0$ and $t=T/2$. The Hamiltonian of a
2D TR-symmetric insulator can easily be put in this context
by treating $k_x$ as the wavevector $k$ of a 1D periodic system while
treating $k_y$ as the pumping parameter $t$. Assuming at the
TR-invariant values of $t$ a gauge of the form
\begin{eqnarray}
\theta |u_{1k}\rangle &=& e^{i\chi_k}|u_{2-k}\rangle \nonumber \\ 
\theta |u_{2k}\rangle &=& -e^{i\chi_{-k}}|u_{1-k}\rangle,
\label{chi}
\end{eqnarray}
that is smooth in $k$, it was shown that one can compute the
$\z2$ invariant associated with the pumping process from a
knowledge of the occupied states at the TR-invariant points
of the pumping cycle only. However, for this purpose
the gauge must be smooth on the whole torus formed by
$k$ and $t$.\cite{Fu-PRB06}

Let us now look at how all this is reformulated in terms
of the gauges introduced in the present paper for a 2D system.
The Hamiltonian gauge of an ordinary TR-symmetric
insulating system corresponds to $\chi_k=0$ in Eq.~(\ref{chi}),
and it is possible to define Bloch states in a smooth
fashion on the whole torus subject to this condition.
However, for a $\z2$ insulator such a constraint introduces a
topological obstruction for a smooth gauge.\cite{Fu-PRB06}
This can be understood in
terms of the cylindrical gauge introduced in Sec.~\ref{sec:cherns}.
Taking into account that the TR-symmetric values of the pumping
parameter now correspond to $k_y=0$ and $k_y=\pm \pi$, note that
in the cylindrical gauge the
TR operator maps the states at $(k_x,k_y=0)$ to
$(-k_x,k_y=0)$ and the states at $(k_x,k_y=\pm \pi)$ to
$(-k_x,k_y=\mp \pi)$ according to Eq.~(\ref{chi0}).
If we now take into account the boundary
conditions of Eq.~(\ref{1band2D}) for the cylindrical gauge and
use them to relate the states at $(k_x,k_y)$ to those at
$(-k_x,k_y)$, one then arrives at a relation of the
form of Eq.~(\ref{chi}) with
$$
\chi_k=0
$$
at $k_y=0$ and
$$
\chi_k=\pm k_x C
$$
at $k_y=\pm \pi$.
For an ordinary insulator $C=0$, and this obviously
reduces to the standard case of $\chi_k=0$ both at $k_y=0$
and $k_y=\pm \pi$.

To derive an expression for the $\z2$ invariant
a concept of partial polarization was
introduced\cite{Fu-PRB06} using the gauge of Eq.~(\ref{chi}) via
\begin{equation}
P^{(S)}_{t}=\frac{1}{2\pi}\left[i\int_{0}^{\pi}\langle u_{S,t,k}|
\partial_k |u_{S,t,k} \rangle dk +(\chi_{t,k=\pi}-\chi_{t,k=0})\right]
\end{equation}
where the index $S=1,2$ differentiates between the two states of
a Kramers pair. This expression is $\mathrm{U(2)}$
invariant modulo a lattice vector ($a=1$), provided that the
transformation is globally smooth in 1D. The $\z2$ invariant
was defined as
\begin{equation}
\nu=(P^{(1)}_{t=0}-P^{(2)}_{t=0})-
(P^{(1)}_{t=T/2}-P^{(2)}_{t=T/2}),
\end{equation}
when the gauge is also smooth in $t$ from $0$ to $T/2$.
With the $\chi_k$ suggested by the cylindrical gauge, and
taking into account that $C$ has opposite sign for $S=1$
and $S=2$, one has
$P^{(1)}_{k_y=0}-P^{(2)}_{k_y=0}=0$ and
$P^{(1)}_{k_y=\pm \pi}-P^{(2)}_{k_y=\pm \pi}=\pm C$, obviously giving the
correct value of the topological invariant.

As shown above, the construction
of a smooth gauge starting with the
cylindrical gauge proceeds by means of a unitary rotation that
unwinds the gauge discontinuity of the cylindrical gauge. The
unitary matrix that realizes this transformation is smooth
and periodic in $k_x$. Thus, when establishing a smooth gauge
at the TR-invariant values of $k_y$ the gauge
condition~(\ref{chi}) on the 1D system is changed smoothly and,
as discussed in Sec.~\ref{sec:rotation},
the smooth occupied subspaces are no longer
mapped onto each other by $\theta$. However,
the TR polarization does not change under such a
transformation, and as was nicely shown in
Ref.~\onlinecite{Fu-PRB06},
one can compute the $\z2$ index using the formula
\begin{equation}
(-1)^{\nu}=\prod_{i=1}^4 \frac{\sqrt{\det[w(\k^*_i)]}}
{\mathrm{Pf}[w(\k^*_i)]},
\label{pfaffian}
\end{equation}
where $\k^*$ are TR-invariant
momenta (i.e., ${\bf k}^*=-{\bf k}^*+{\bf G}$) and
\begin{equation}
w_{ij}({\bf k})=\langle u_{i-{\bf k}}|\theta
|u_{j{\bf k}}\rangle.
\label{wmat}
\end{equation}
Note, that $w(-{\bf k}^*)=-w^T({\bf k}^*)$, so that
the Pfaffian in~(\ref{pfaffian}) is well defined.

The application of the smooth-gauge construction developed in the
main text of this paper to the Kane-Mele model in the QSH regime
indeed results in the odd value for $\nu$.  The TR constraint
takes the form $w_{12}(\k^*)=\pm 1$ and, as mentioned
above, is satisfied only at the TR-invariant momenta, with 
$|w_{12}(\k)|<1$ at other values of $\k$. In
particular, our parameter choice ($\lambda_v/t=1$, $\lambda_{SO}/t=0.6$, 
$\lambda_{R}/t=0.5$) results in $w_{12}(0,0)=1$ but
$w_{12}(0,\pi)=w_{12}(\pi,0)=w_{12}(\pi,\pi)=-1$, thus signaling
a band inversion at $\Gamma=(0,0)$.

\bibliography{paper}
\end{document}